\documentstyle[pre,aps,epsfig,citesort,amsmath]{revtex}

\begin{document}
\title{Homogeneous cooling of rough, dissipative particles:
       Theory and simulations}

\author{S. Luding$^{(1)}$, 
        M. Huthmann$^{(2)}$, S. McNamara$^{(1,3)}$, A. Zippelius$^{(2)}$ }
\address{(1) Institute for Computer Applications 1, Pfaffenwaldring 27, 
70569 Stuttgart, GERMANY\\
(2) Institut f\"ur Theoretische Physik, Universit\"at G\"ottingen, 
Bunsenstr. 9, 37073 G\"ottingen, GERMANY\\
(3) Levich Institute, Steinman Hall T-1M, 140th St and Convent Ave,
New York, NY 10031, USA}
\date{\today}
\twocolumn
\maketitle

\begin{abstract}
We investigate freely cooling systems of rough spheres in two 
and three dimensions. Simulations using an event driven algorithm 
are compared with results of an approximate kinetic
theory, based on the assumption of a generalized homogeneous cooling state.
For short times $t$, translational and rotational energy are found to 
change linearly with $t$. For large times both energies decay
like $t^{-2}$ with a ratio independent of time, but {\em not}
corresponding to equipartition. Good agreement is found between theory and 
simulations, as long as no clustering instability is observed. 
System parameters, i.e. density, particle size, and particle mass 
can be absorbed in a rescaled time, so that the decay of translational
and rotational energy is solely determined by normal restitution and
surface roughness.\\

\end{abstract}

\section{Introduction}

Collections of macroscopic, dissipative, and rough constituents have
attracted a lot of interest recently, mainly in the context of
granular media \cite{herrmann98,goldhirsch93,mcnamara96}. 
In the so-called grain-inertia regime, one is concerned with rapid
granular flow, which on one hand can be described by methods of
statistical mechanics, analogous to the kinetic theory of dense gases
\cite{jenkins85b,lun87,lun91,goldshtein95,huthmann97}, and,
on the other hand, can be well simulated with help of event-driven 
algorithms \cite{luding95b,mcnamara98}.
In both approaches the dynamics of the system is assumed to be
dominated by two-particle collisions which are modelled by their
asymptotic states. A collision is characterized by the velocities
before and after the contact, and the contact is assumed to be 
instantaneous. In the simplest model, one describes inelastic
collisions by normal restitution only. However, surface roughness is
important for the dynamics of granular flow 
\cite{luding95b,huthmann97,mcnamara98} and allows for an exchange
of translational and rotational energy.

The rough sphere model was first introduced for elastically colliding
particles, assuming either perfectly rough or perfectly smooth
particles \cite{bryan1894,chapman60}.  Subsequently, it has been
generalised to intermediate values of the roughness to account for
tangential restitution in inelastic collisions. Several groups have
investigated the exchange of translational and rotational energy,
using kinetic theory
\cite{jenkins85b,lun87,lun91,goldshtein95,huthmann97} or numerical
simulations \cite{luding95b,mcnamara98}.  One major result is that in
contrast to conservative systems, energy is not equipartitioned
between the degrees of freedom in dissipative systems
\cite{luding95b,huthmann97,mcnamara98,jenkins85b,lun87,lun91,goldshtein95},
even though the ratio of translational and rotational energy
approaches a constant, while both functions decay to zero in a freely
cooling system.  In this paper we study the full time evolution of
translational and rotational energy of a gas of rough, hard spheres.
We compare numerical simulations to kinetic theory, which is based on
a pseudo--Liouville--operator formalism and makes use of the
assumption of a homogeneous state. The computation covers the full
range of times, from the initial linear change of energies with time,
to the asymptotic state for large times, and also including the
crossover between the two
limiting regimes.\\

In section\ \ref{microdyn} we introduce the microscopic interaction laws
and the operators needed in section\ \ref{homcool} to derive the solution
for homogeneously cooling systems. The analytical solution is compared
to the simulations in section\ \ref{numeric} and the results are 
summarized and discussed in\ \ref{discuss}.

\section{Microscopic Dynamics}
\label{microdyn}

We consider a system of $N$ $D$-dimensional spheres interacting via a
hard-core potential confined to a $D$-dimensional volume $V = L^D$
with linear size $L$ and $D = 2$ or $3$. The
spheres have mass $m$, radius $a$, and moment of inertia $I$. The
degrees of freedom are the positions $\mbox{\boldmath$r$}_{\mu}(t)$,
the translational velocities $\mbox{\boldmath$v$}_{\mu}(t)$,
and the angular velocities $\mbox{\boldmath$\omega$}_{\mu}(t)$ for each
sphere, numbered by $\mu=1,2,\ldots,N$. Here, the spheres are 
rough with constant normal and also constant tangential restitution. 
The translational and angular velocities after collision (primed quantities) 
are determined by the velocities before collision (unprimed quantities)
so that 
\begin{eqnarray}
 \mbox{\boldmath$v$}_{\mu}' & = &
         \mbox{\boldmath$v$}_{\mu}-\frac{1+r}{2}\mbox{\boldmath$v$}_n
         -\frac{q(1+\beta)}{2q+2}
                             (\mbox{\boldmath$v$}_t+\mbox{\boldmath$v$}_r)~,
   \label{eq:colla} \\ 
\mbox{\boldmath$v$}_{\nu}' & = &
          \mbox{\boldmath$v$}_{\nu}+\frac{1+r}{2}\mbox{\boldmath$v$}_n
         +\frac{q(1+\beta)}{2q+2}
                              (\mbox{\boldmath$v$}_t+\mbox{\boldmath$v$}_r)~,
   \label{eq:collb} \\ 
\mbox{\boldmath$\omega$}_{\mu}' & = &
         \mbox{\boldmath$\omega$}_{\mu} 
         + \frac{1+\beta}{a(2q+2)}
                [\hat{\mbox{\boldmath$r$}}\times(\mbox{\boldmath$v$}_t + 
                                               \mbox{\boldmath$v$}_r) ]~, 
     {\rm ~~and} \label{eq:collc} \\ 
\mbox{\boldmath$\omega$}_{\nu}' & = &
         \mbox{\boldmath$\omega$}_{\nu} 
         +\frac{1+\beta}{a(2q+2)}
                [\hat{\mbox{\boldmath$r$}}\times(\mbox{\boldmath$v$}_t + 
                                                  \mbox{\boldmath$v$}_r) ] ~.
   \label{eq:colld} 
\end{eqnarray}
The unit vector
$\hat{\mbox{\boldmath$r$}}=(\mbox{\boldmath$r$}_\mu-\mbox{\boldmath$r$}_\nu)/
|\mbox{\boldmath$r$}_\mu-\mbox{\boldmath$r$}_\nu|$ points from particle
$\nu$ to $\mu$, along the line connecting the centres of mass. 
We have introduced abbreviations for the normal velocity 
$\mbox{\boldmath$v$}_n=[(\mbox{\boldmath$v$}_\mu-\mbox{\boldmath$v$}_\nu)
\cdot\hat{\mbox{\boldmath$r$}})]\cdot \hat{\mbox{\boldmath$r$}}$,
for the tangential velocity due to translational motion
$\mbox{\boldmath$v$}_t=\mbox{\boldmath$v$}_\mu-
\mbox{\boldmath$v$}_\nu-\mbox{\boldmath$v$}_n$
and for the tangential velocity due to rotational motion
$\mbox{\boldmath$v$}_r=-a(\mbox{\boldmath$\omega$}_\mu
+\mbox{\boldmath$\omega$}_\nu)\times\hat{\mbox{\boldmath$r$}}$. 
The constants $r$ and $\beta$ characterize normal and tangential
restitution, and $q={I}/({ma^2})$ depends on the mass distribution
inside a grain.\\

The time evolution of a dynamical variable
$A(t)=A(\Gamma;t)=A(\{\mbox{\boldmath$r$}_\mu(t),\mbox{\boldmath$v$}_\mu(t),
\mbox{\boldmath$\omega$}_\mu(t)\};t)$
is (for positive times) determined by the pseudo--Liouville--operator 
${\cal L}_{+}$
 \begin{equation}
A(t) = \exp( i {\cal L}_{+} t) A(0)~.
 \end{equation}
The pseudo--Liouville--operator \cite{ernst69,noije97,huthmann97}
 consists of two
parts ${\cal L}_{+}={\cal L}_0 + {\cal L}^{'}_{+}$. The first,
${\cal L}_0$, describes the undisturbed motion of single particles
 \begin{equation}
   {\cal L}_0 = 
           -i \sum_\mu \mbox{\boldmath$v$}_\mu \cdot 
                               {\nabla}_{\mbox{\boldmath$\scriptstyle r$}_\mu}
 \end{equation}
and the second, ${\cal L}^{'}_{+}=\frac{1}{2}\sum_{\mu \ne \nu}{\cal C}_{+}(\mu\nu)$, 
describes hard-core collisions of all pairs of particles ($\mu$, $\nu$) with
 \begin{multline} \label{stossit}
   {\cal C}_{+}(\mu\nu) = \\
   i(\mbox{\boldmath$v$}_{\mu\nu}\cdot\hat{\mbox{\boldmath$r$}}_{\mu\nu}) 
   \Theta(-\mbox{\boldmath$v$}_{\mu\nu}\cdot\hat{\mbox{\boldmath$r$}}_{\mu\nu} )
   \delta(|\mbox{\boldmath$r$}_{\mu\nu}|- 2 a)(b_{\mu\nu}^{+}-1)~. 
 \end{multline}
The operator $b_{\mu\nu}^+$ acts on the particles $\mu$ and $\nu$
only, and instantaneously replaces the translational and angular
momenta just before the collision, by the corresponding values just
after.  
$\Theta
(-\mbox{\boldmath$v$}_{\mu\nu}\cdot\hat{\mbox{\boldmath$r$}}_{\mu\nu})$ is the Heaviside step--function which is non-zero for
approaching particles only. The term
$\mbox{\boldmath$v$}_{\mu\nu}\cdot\hat{\mbox{\boldmath$r$}}_{\mu\nu}$
gives the relative velocity of two colliding particles, reflecting
that fast particles collide more frequently.  Finally, we have
introduced the notation
$\mbox{\boldmath$v$}_{\mu\nu}=\mbox{\boldmath$v$}_\mu-\mbox{\boldmath$v$}_\nu$,
$\mbox{\boldmath$r$}_{\mu\nu}=\mbox{\boldmath$r$}_\mu-\mbox{\boldmath$r$}_\nu$,
and
$\hat{\mbox{\boldmath$r$}}_{\mu\nu}=\mbox{\boldmath$r$}_{\mu\nu}/|\mbox{\boldmath$r$}_{\mu\nu}|$.

\section{Homogeneous cooling state}
\label{homcool}
The ensemble average of a dynamical variable $A(\Gamma;t)$ is defined by
\begin{equation}
\label{eq:A_t}
   \left <A \right >_t  =
    \int d\Gamma \, \rho (\Gamma;0) A(\Gamma;t)= \int d\Gamma \, \rho
    (\Gamma;t) A(\Gamma;0)  
\end{equation}
with the abbreviation for phase space integration 
\begin{equation}
\int d\Gamma = \int \prod_\mu(d\mbox{\boldmath$r$}_\mu d\mbox{\boldmath$v$}_\mu 
                d\mbox{\boldmath$\omega$}_\mu)
          \prod_{\mu<\kappa}\Theta(|\mbox{\boldmath$r$}_{\mu\kappa}|-2 a) ~.
\label{eq:dgamma}
\end{equation}
Here $\rho(\Gamma;t)$ is the $N$-particle phase space distribution function,
i.e.\ $\rho(\Gamma;t)\,d\Gamma$ is the probability at time $t$ to find 
particle 1 at 
$(\mbox{\boldmath$r$}_1,\mbox{\boldmath$v$}_1,\mbox{\boldmath$\omega$}_1)$,
particle 2 at 
$(\mbox{\boldmath$r$}_2,\mbox{\boldmath$v$}_2,\mbox{\boldmath$\omega$}_2)$,
etc.\ The time evolution of the $N$-particle distribution 
$\rho(\Gamma;t)=\exp {(-i{\cal L}_+^{\dagger}t)}\,\rho(\Gamma;0)$ is
governed by the adjoint ${\cal L}_+^{\dagger}$ of the time evolution operator
${\cal L}_+$. 

The quantities of interest are the translational and
rotational energies per particle 
$E_{\rm tr}  = {m}/({2N}) \sum_\mu  \mbox{\boldmath$v$}_\mu^2$ and
$E_{\rm rot} = {I}/({2N}) \sum_\mu  \mbox{\boldmath$\omega$}_\mu^2$,
as well as the total kinetic energy $E=E_{\rm tr}+E_{\rm rot}$. It is
impossible to calculate these expectation values exactly. To make some
progress we resort to an approximation, known as homogeneous cooling
state. One assumes that the $N$-particle distribution function is
spatially homogeneous and depends on time only via the average kinetic energy
of the grains. It was shown in \cite{huthmann97} that initially
equipartitioned translational and rotational energy change with different 
rates, suggesting a generalised cooling state of the form
\begin{equation}
\label{rohcs}
  \rho_{\rm HCS}(\Gamma;t) \sim \exp\left[ - N \left( 
       \frac{E_{\rm tr}}{T_{\rm tr}(t)}+ \frac{E_{\rm rot}}{T_{\rm rot}(t)}
                    \right )\right]~, 
\end{equation}
The $N$-particle distribution does not depend on the spatial
coordinates $\{\mbox{\boldmath$r$}_\mu\}$ as a consequence of the
assumption of homogeneity. It depends on time only via the average
translational and rotational energy:
$T_{\rm tr}(t)={2/D} \left <E_{\rm tr} \right >_t $ and 
$T_{\rm rot}(t)=2/({2D-3}) \left <E_{\rm rot} \right >_t $. The factors 
$D$ and $2D-3$ account for the number of translational and rotational 
degrees of freedom in $D$ dimensions respectively. 
To study the time evolution of the average translational and
rotational energy we consider the corresponding time derivatives:

\begin{align} \nonumber
  \frac{d}{dt}  T_{\rm tr}(t)  & = 
  \frac{2}{D} \int d\Gamma \,\rho(\Gamma;t) i {\cal L}_+ {E_{\rm tr}}
  ~{\rm~~and} \\ 
  \frac{d}{dt}  T_{\rm rot}(t) & =  
  \frac{2}{2D-3}\int d\Gamma \,\rho(\Gamma;t) i {\cal L}_+ {E_{\rm rot}} ~.
  \label{derivative}
\end{align}
Then, we assume the generalized homogeneous cooling state of Eq.\ (\ref{rohcs}),
replacing $\rho(\Gamma;t)$ by $\rho_{\rm HCS}(\Gamma;t)$ in Eqs.\ (\ref{derivative}), 
and arrive at 
\begin{align} \nonumber
  \frac{d}{dt}  T_{\rm tr}(t)  & 
   = \frac{2}{D} \left\langle i {\cal L}_+ {E_{\rm tr}} \right\rangle_{\rm HCS} 
  ~{\rm~~and} \\ 
  \frac{d}{dt}  T_{\rm rot}(t) & 
   = \frac{2}{2D-3} \left\langle i {\cal L}_+ {E_{\rm rot}} 
                                                      \right\rangle_{\rm HCS}~.
  \label{eq:hcs}
\end{align}
All that remains to be done is a high dimensional phase space
integral, the details of which are delagated to appendix A, where we
present calculations for $D=2$.
After integration over phase space has been performed, we find
\begin{align}\nonumber
  \left\langle i {\cal L}_+ {E_{\rm tr}} \right\rangle_{\rm HCS} & = 
  - G A T_{\rm tr}^{3/2} + G B T_{\rm tr}^{1/2} T_{\rm rot} 
  ~{\rm~~and} \\ 
  \left\langle i {\cal L}_+ {E_{\rm rot}} \right\rangle_{\rm HCS} & = 
  G B T_{\rm tr}^{3/2} - G C T_{\rm tr}^{1/2} T_{\rm rot} 
  \label{eq:hcs2}
\end{align}
with the constants $A$, $B$, $C$, and $G$, whose values depend on space
dimensionality $D$.

\subsection{The 2D-Case}
In two dimensions the constants in Eqs.\ (\ref{eq:hcs2}) are given by
\begin{align} \nonumber
G & = 8 a \frac{N}{V}\sqrt{\frac{\pi}{m}} g(2a), &\quad
A & =\frac{1-r^2}{4}+\frac{\eta}{2}(1-\eta),            \\ 
B & = \frac{\eta^2}{2q},                      &\quad 
C & = \frac{\eta}{2q} \left (1-\frac{\eta}{q} \right ). 
\label{eq:const2d}
\end{align}
We have introduced the abbreviation 
$\eta = {q(1+\beta)}/{(2q+2)}$, and $g(2 a)$ 
denotes the pair correlation function 
at contact. We consider homogeneous discs with $q=1/2$.

In a system of elastically colliding particles \cite{chapman60},
$G\, T_{\rm tr}^{1/2}$
is just the Enskog collision frequency, whose inverse determines the
mean time between collisions. We use this mean collision frequency of
the elastic system to 
rescale time according to $\tau = G\, T_{\rm tr}^{1/2}(0)\, t$, and 
furthermore introduce the dimensionless translational temperature 
$T = T_{\rm tr}(t)/T_{\rm tr}(0)$ and the dimensionless
rotational temperature $R = T_{\rm rot}(t)/T_{\rm tr}(0)$. 
Expressed in scaled units Eqs.\ (\ref{eq:hcs}) and (\ref{eq:hcs2})
read in $2D$:
\begin{align}\nonumber
\frac{d}{d \tau} T & = -A T^{3/2} + B T^{1/2}R,~\\
\frac{d}{d \tau} R & = 2 B T^{3/2} - 2 C T^{1/2}R~.
\label{eq:TR}
\end{align}
To find the asymptotic value of the ratio of $K=T/R$ we consider the
differential equation 
\begin{equation}
 \frac{dT}{dR}= \frac{-A K + B}{2B K- 2C}~,
\end{equation}
which can be solved by a constant 
\begin{equation}
K = \left(2 C- A + \sqrt{(2C-A)^2+8B^2}\right)/(4B)  
\label{eq:ratio}
\end{equation}
in agreement with Ref.\ \cite{mcnamara98}.  

\begin{figure}[htb]
\begin{center}
\epsfig{file=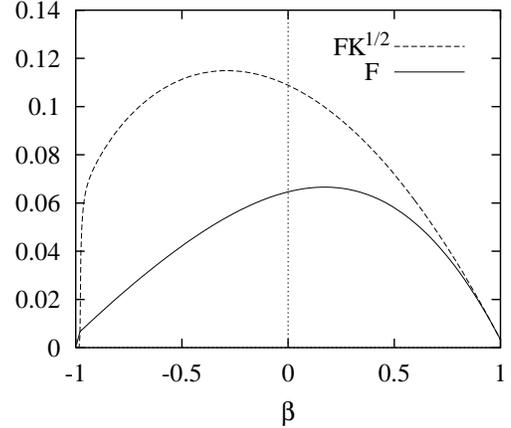,width=6.2cm,angle=-90} 
\end{center}
\caption{Dependence of the asymptotic decay-prefactors
         $F$ and $F \sqrt{K}$  on $\beta$ for $r=0.99$.
}
\label{fig:scafig1}
\end{figure}

Of particular interest is the long time decay of translational and
rotational energy as a function of $r$ and $\beta$. We would expect
the decay to be slowest, if energy is lost only due to normal
restitution and not due to roughness. Hence, as a function of $\beta$,
translational and
rotational energy should persist for the longest times for $\beta =\pm 1$.
In between, i.e. $-1<\beta <1$, we expect a faster decay, because
surface roughness is an additional mechanism for the disspation of energy.
These expectations are born out by the following detailed discussion
of Eqs.(\ref{eq:TR},\ref{eq:ratio}). We assume that the ratio $K = T/R $ has
reached its asymptotic value given in Eq.\ (\ref{eq:ratio}) for some 
$\tau > \tau_0$ and substitute $R=T/K$ into Eq.\ (\ref{eq:TR})
and obtain
\begin{equation}
\frac{d}{d \tau} T  = - F T^{3/2}.
\end{equation}
The resulting equation is of the same functional form as for
homogeneous cooling of smooth spheres, except for the coefficient 
$F=A-B/K$, which contains all the dependence on $\beta$ and $r$.
Its solution is given by
\begin{equation}
  T = \frac{T(\tau_0)}{\left[1+T(\tau_0)^{1/2}(F/2)(\tau-\tau_0)\right]^2}~.
\end{equation}
Asymptotically, for $\tau \rightarrow \infty$, the translational energy decays 
like $T \approx (\frac{F}{2}\tau)^{-2}$ and the rotational energy like 
$R \approx (\frac{F\sqrt{K}}{2}\tau)^{-2}$.
In a double-logarithmic plot of $T$ and $R$ against $\tau$ one should
observe straight lines with slope $-2$ and axial sections (at $\tau=1$) 
given by $-2\ln{(F/2)}$ for the translational energy, and by 
$-2\ln{(F\sqrt{K}/2)}$ for the rotational energy. For the same slope a
larger axial section implies persistence for longer times, i.e. a 
`slower' decay in time.
In Fig.\ \ref{fig:scafig1} we plot the prefactors $F$ and $F\sqrt{K}$ 
against $\beta$ for $r=0.99$. 

As a function of $\beta$, $F$ is smallest for
$\beta\rightarrow \pm 1$ corresponding to the cases
where no energy is lost due to friction. The decay of translational
energy is then pushed out to the longest timescales. 
In Fig.\ \ref{fig:scafig1}, the maximum of
$F$ is reached for $\beta_{\rm tr}^{\rm max} \approx 0.17$, 
corresponding to the `fastest' decay of $T$. 
For $\beta = -1$ we find that $F \sqrt{K} = 0$,
because the rotational energy remains constant as a function of time.
 $F \sqrt{K}$ reaches its maximum 
for $\beta_{\rm rot}^{\rm max} \approx -0.29 \neq \beta_{\rm tr}^{\rm
  max}$, 
so that the decay of 
rotational energy is `fastest' for a value of $\beta$, different 
from the one where the translational energy decays fastest.
In the limit $\beta \rightarrow 1$ the axial sections for $T$ and $R$
have approximately the same values. In this case the ratio $K$ is close to
unity, reflecting the rather effective exchange of rotational and translational 
energy for rough spheres and thus approximate equipartition.

\subsection{The 3D-Case}
In 3D we consider spheres with $q=2/5$. 
The constants in Eq.\ (\ref{eq:hcs2}) are given by 
\begin{align} \nonumber
G & = 8 (2a)^2 \frac{N}{V}\sqrt{\frac{\pi}{m}} g(2a), &\quad
A & = \frac{1-r^2}{4}+\eta (1-\eta),            \\ 
B & = \frac{\eta^2}{q},                      &\quad 
C & = \frac{\eta}{q} \left (1-\frac{\eta}{q} \right )~. 
\label{eq:const3d}
\end{align}
As in two dimensions we introduce a dimensionless time 
$\tau = \frac{2}{3} G T_{\rm tr}^{1/2}(0)\,t$ with the
factor $2/D=2/3$, which accounts for the number of translational 
degrees of freedom. The time dependence of the
dimensionless translational temperature $T = T_{\rm  tr}(t)/T_{\rm tr}(0)$ 
and of the dimensionless rotational temperature 
$R =T_{\rm rot}(t)/T_{\rm tr}(0)$ follows from
\begin{align}\nonumber
\frac{d}{d \tau} T & = -A T^{3/2} + B T^{1/2}R~,\\
\frac{d}{d \tau} R & =  B T^{3/2} - C T^{1/2}R~.
\label{eq:TR2}
\end{align}
The asymptotic value of the ratio of $K=T/R$ is given in Refs.\ 
\cite{goldshtein95} and \cite{mcnamara98}:
\begin{equation}
K = \left(C- A + \sqrt{(C-A)^2+4B^2}\right)/(2B)~.  
\end{equation}
The long time limit can be discussed as in the 2D-case. We find a
very similar result for the dependence of the asymptotic decay 
on $r$ and $\beta$.

The full time dependent
solution, obtained by numerical integration of Eqs.\ (\ref{eq:TR}) 
and (\ref{eq:TR2}), will be compared with simulations in the next section.

\section{Numerical Experiments}
\label{numeric}
Since we are interested in the behavior of granular particles
cooling over several decades in time, we use an event driven (ED) method.
In these simulations, the particles follow an undisturbed translational
motion until an event occurs. An event is either the collision
of two particles or the collision of one particle with a boundary
of a cell (in the linked-cell structure). 
The cells have no effect on the particle-motion here; 
they were solely introduced to accelerate the search for future 
collision partners in the algorithm (see also Appendix B). 
Using the velocities just before contact we compute the particles' 
velocities just after the contact following 
Eqs.\ (\ref{eq:colla}-\ref{eq:colld}). 
In the ED method the contact duration is implicitly zero, matching 
well the corresponding assumption of instantaneous contacts used for 
the theory.
We remark that ED algorithms run into problems when the time between
events $t_{\rm ev}$ gets too small. In dense systems with
strong dissipation $t_{\rm ev}$ may tend towards zero. As
a consequence the so-called ``inelastic collapse'' can
occur, i.e. the divergence of the number of events per unit time.
The problem of the inelastic collapse \cite{mazighi94,mcnamara94},
can be handled using restitution coefficients dependent on the 
time elapsed since the last event \cite{luding96e,luding97c}.
For the contact that occurs at time $t_{ij}$, one uses $r=1$ and 
$\beta=-1$ if at least one of the involved partners had a collision 
with another particle later than $t_{ij}-t_{\rm c}$. The time $t_{\rm c}$
can be identified as a typical duration of a contact.
The effect of $t_{\rm c}$ on the simulation
results is negligible for large $r$ and small $t_{\rm c}$, 
what we checked in the simulations. Only for very small values
$r \le 0.6$ a cut-off time $t_{\rm c} = 10^{-6}$\,s was needed to avoid
the inelastic collapse. For a more detailed discussion of the ED 
algorithm used, see Appendix B.

Every simulation is first equilibrated with $r=1$ and $\beta=-1$
until the velocity distribution is Maxwellian. Then the
restitution coefficients are set to the selected values.
In order to classify the systems used for the simulations
we need to specify the number of particles $N$ and the volume
fractions $\rho_{\rm 2D} = (N/V) \pi a^2$ and
$\rho_{\rm 3D} = (N/V) (4/3) \pi a^3$ in two and three dimensions,
respectively.
 
\subsection{Results in 2D}

For short times we can solve Eqs.\ (\ref{eq:TR}) analytically and
get $T = 1 - A \tau $ and $R = (2 B) \tau $. Hence our data can be
effectively collapsed for short times by rescaling
time accorrding to $\tau \rightarrow A \tau$
and rotational temperature according to $R\rightarrow RA/(2B)$.
We compare the theoretical result [numerical solution of Eq.\ (\ref{eq:TR})] 
with simulations for various sets of parameters. The scaling
collapses data for different particle number, particle radius and density
on the same master curves, since all these dependencies are hidden in
the rescaled time $\tau$. The shape of the master curves depends only
on the restitution coefficients $r$ and $\beta$.
In order to check the dependence of the solution on 
system size we simulate a {\em large} system with $N=99856$ particles 
and a {\em small} system with only $N=198$. For the small system we
consider two volume fractions
$\rho=0.25$ or $\rho=0.01$, the latter corresponding to an extremely 
{\em dilute} system. For the large system we always use $\rho=0.25$.
For the larger volume fraction we get numerically
$g_{0.25}(2a) \approx 1.58$ in perfect agreement with the 2D 
equivalent of the Carnahan-Stirling formula
\begin{equation}
 g(2a) = \frac{1 - 7 \rho / 16}{(1-\rho)^2}~,
\end{equation}
from Ref.\ \cite{verlet82}.
Initially, the normalized energies are $T=1$ 
and $R=0$ for all data presented here. 

\begin{figure}[htb]
\begin{center}
\epsfig{file=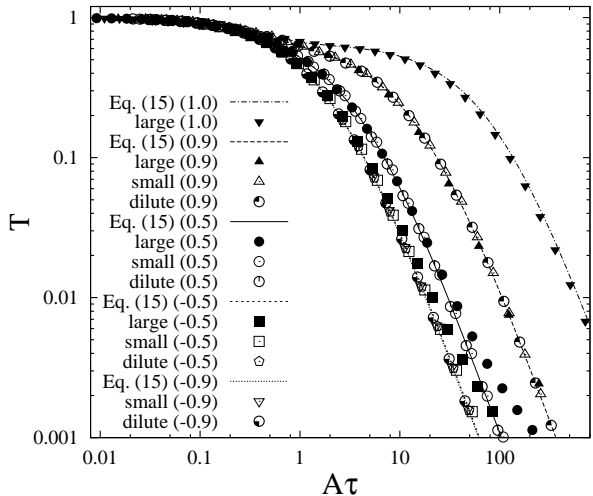,height=8.5cm,angle=-90} 
\end{center}
\caption{
$T$ as function of rescaled time $A \tau$. Different symbols
correspond to different simulations with $N=99856$, $\rho=0.25$
(large), $N=198$, $\rho=0.25$ (small), and $N=198$, $\rho=0.01$
(dilute). The coefficient of restitution is $r=0.99$ and the
value of $\beta$ is given in brackets in the inset.
The curves represent numerical solutions of Eqs.\ 
(\protect{\ref{eq:TR}}). 
}
\label{fig:fig1}
\end{figure}

In Fig.\ \ref{fig:fig1} we plot $T$ 
against rescaled time $A \tau$ for various simulations
with fixed value of $r=0.99$ and 
different particle number $N$, volume fraction $\rho$,
and tangential restitution $\beta$. We observe that all
simulations with the same $\beta$ collapse, independent
of the specific value of $N$ or $\rho$. This indicates that
the dimensionless units $T$ and $\tau$ are indeed the 
system's inherent ``temperature'' and ``time''. The temperature
$T$ decays continuously from its initial value $T=1$, following
the function $T=1-A \tau$ for short times $A \tau < 0.1$ and decaying like
$\tau^{-2}$ for long times. 
The monotonic dependence of $T$ on $\beta$ is due to our scaling of
the horizontal axis, which includes factors of 
$\beta$ via $A$. If we plot $T$ against
$\tau$ we observe that the crossover between the initial linear decay
and the asymptotic $\tau^{-2}$ behavior is nonmonotonic with $\beta$: The
time of crossover is largest for $\beta =-1$, then decreases and
reaches a minimum around $\beta_{\rm tr}^{\rm max} \approx 0.17$ and 
then increases again
for $\beta \rightarrow 1$ [in agreement with Fig.\ref{fig:scafig1}].

The simulations of the {\em large} system deviate from the 
theory and also from the simulations for the small system
at large times or small $T$. The deviation from the theoretical curve is
due to the density instability, i.e.~clusters of particles
form and the homogeneous cooling assumption fails.
Additional simulations show that the deviation from theory 
occurs earlier for stronger dissipation due to either 
smaller $r$ or $|\beta|$.

\begin{figure}[htb]
\begin{center}
\epsfig{file=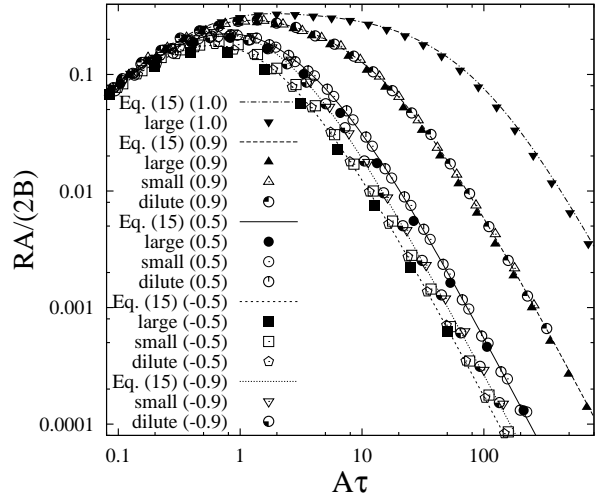,height=8.5cm,angle=-90}
\end{center}
\caption{
$RA/(2B)$ as function of rescaled time $A \tau$. The simulations
are the same as in Fig.\ \protect{\ref{fig:fig1}}.
The curves represent the numerical solutions of 
Eqs.\ (\protect{\ref{eq:TR}}). 
}
\label{fig:fig2}
\end{figure}

In Fig.\ \ref{fig:fig2} we plot $RA/(2B)$ against  normalized time $A
\tau$ for the same set of parameters as in Fig.\ \ref{fig:fig1}. As
expected from the solution for small $A \tau$, we find that $RA/(2B)$
is proportional to $A \tau$ for small times $A \tau < 0.1$. In this
regime, the initially `cold' rotational degrees of freedom are activated
due to the transfer of linear to angular momentum during collisions.
After some equilibrium between translational and rotational
temperature is achieved, both degrees of freedom loose energy in the
long time limit. Like the translational temperature, also the
rotational temperature is independent of $N$ and $\rho$, only $r$ and
$\beta$ determine the decay of rotational energy. 
As for the translational energy, the observed, almost monotonic dependence 
of the crossover time for the rotational energy on $\beta$ is due to our
scaled units $\tau \rightarrow A \tau$ and $R\rightarrow R A/(2B)$. 
If we plot $R$ against $\tau$ we find a similar nonmonotonic
dependence of the crossover time on $\beta$ as for the translational 
energy with
however the minimum of rotational crossover time occuring at a
different value of $\beta_{\rm rot}^{\rm max} \approx -0.29$ 
[see Fig.\ \ref{fig:scafig1}].

The results of simulations
are in very good agreement with the theoretical results, which are
based on the generalized homogeneous cooling assumption for the
$N$-particle distribution function. Note that the agreement extends over
many orders of magnitude in time and covers the initial increase of
$R$, the crossover and the algebraic decay for long times. 

\begin{figure}[htb]
\begin{center}
\epsfig{file=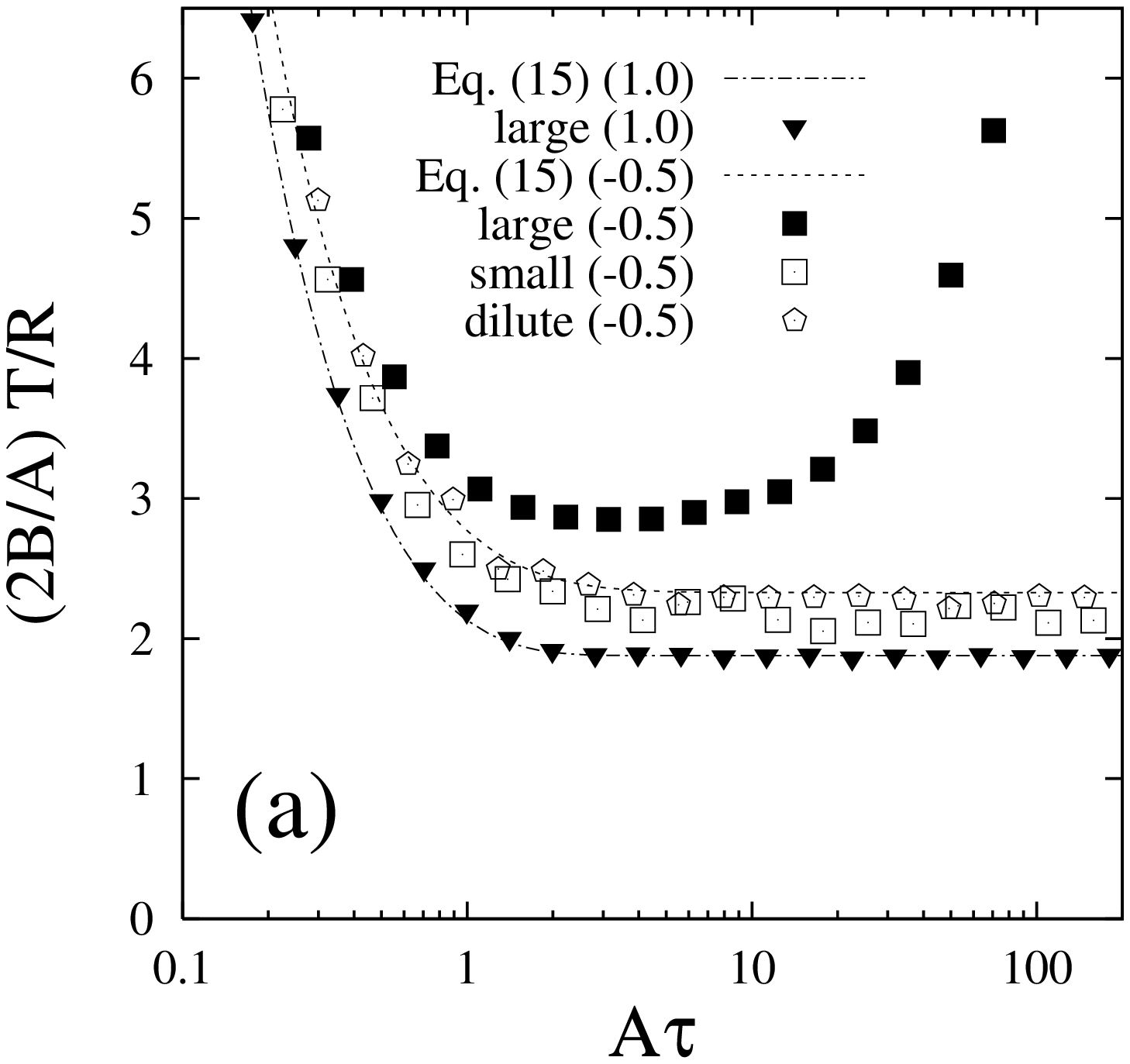,height=6.8cm,angle=-90}
\epsfig{file=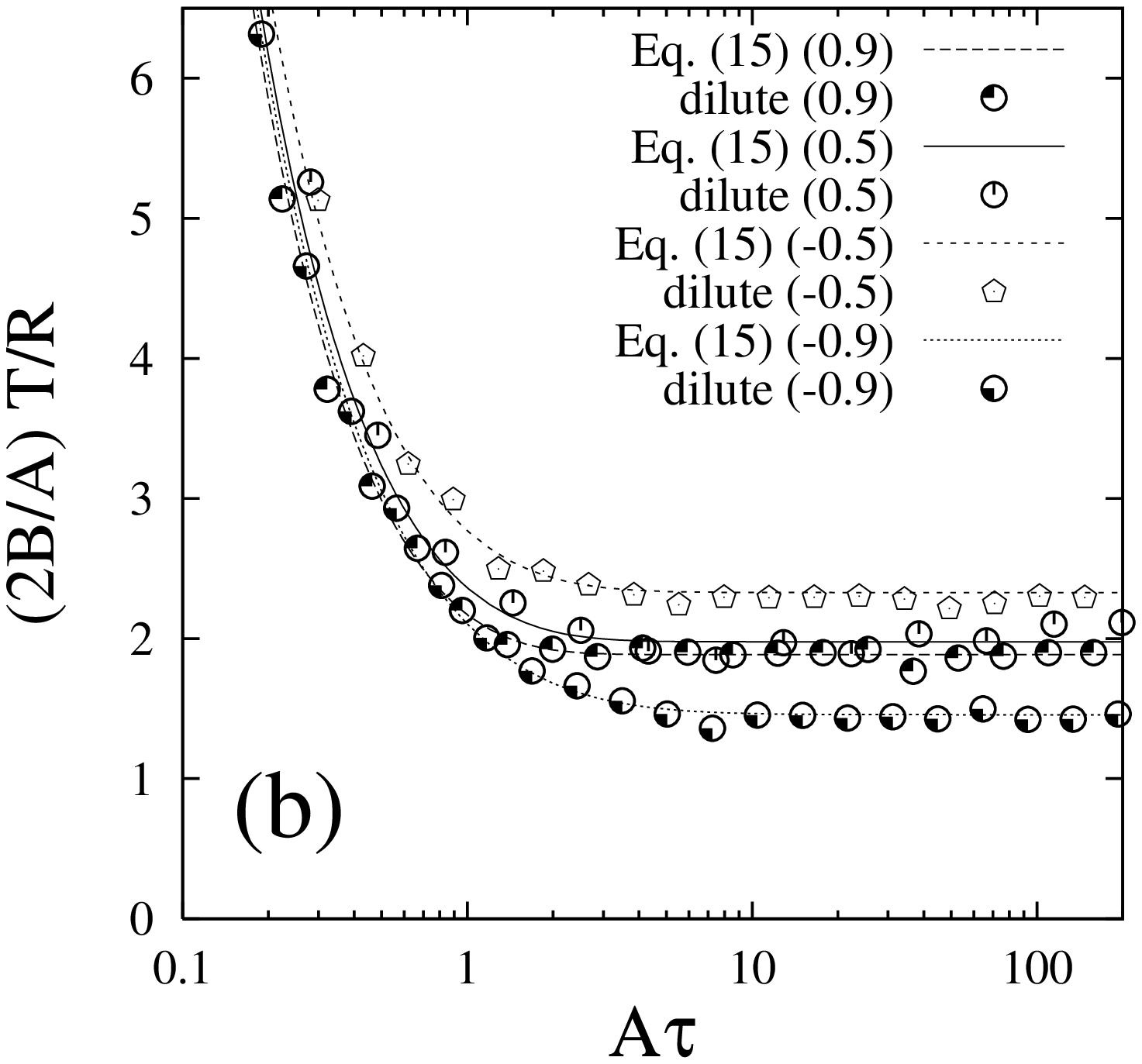,height=6.8cm,angle=-90}
\end{center}
\caption{
$(2B/A) T/R$ as a function of rescaled time $A \tau$ for some
simulations from Fig.\ \protect{\ref{fig:fig1}}.
(a) The simulation of the large system with $N=99856$, $\rho=0.25$, 
and $\beta=-0.5$ (solid squares) shows no saturation of 
$(2B/A)~T/R$ which instead diverges for large $A \tau$.
(b) Here only simulations of the dilute system with $N=198$, and $\rho=0.01$
are compared to theory. All show saturation of the ratio
$(2B/A)~T/R$ for large times. }
\label{fig:fig3}
\end{figure}

In Fig.\ \ref{fig:fig3} we present the ratio of $T$ and
$RA/(2B)$ as a function of scaled time $A \tau$ for a
selected set of parameters which were used in the simulations shown in
 Fig.\ \ref{fig:fig1}. Data for two values $\beta=1.0$ and
 $\beta=-0.5$ are shown in Fig.\ \ref{fig:fig3}(a). For the latter we compare
systems at different densities and observe that the dilute simulations
are in perfect agreement with the theory, whereas the 
dense system with $\rho=0.25$ deviates. The small system
shows a slightly smaller ratio $(2B/A)~T/R$, whereas for the large
system, the ratio $(2B/A)~T/R$ is larger than expected and
eventually diverges for large $A \tau$. This is again the
regime where the homogeneous cooling assumption fails.
Interestingly, the large simulation with $N=99856$, $\rho=0.25$,
and $\beta=1.0$ is in reasonable agreement with theory.
From Fig.\ \ref{fig:fig3}(b),
we learn that the simulations of dilute systems are always in good
agreement with theory. The ratio $(2B/A)~T/R$ is constant
for large times, but there is no equipartition of energies
in the translational and rotational degrees of freedom.

\subsection{Results in 3D}
As in 2D, we can solve Eq.\ (\ref{eq:TR2}) analytically for short times,
and get $T = 1 - A \tau$ and $R = B \tau $.
Hence we rescale time $\tau\rightarrow A\tau$ and 
rotational temperature $R\rightarrow RA/B$.

We simulated various systems characterized by volume fraction $\rho$ and 
particle number $N$. One system with density $\rho=0.087$ and 
$N=1331$ particles is denoted as {\it medium}; two other systems with 
lower density have the parameters $\rho=0.0021$, $N=4096$ {\it dilute},
and $\rho=0.0023$, $N=68921$ {\it large}. The abbreviation corresponds to
the density, only for the large system one should read ``dilute and large''.
Finally, a system with higher density, i.e.~$\rho=0.23$ and $N=54872$ 
{\it dense} is examined. To calculate the pair correlation function at 
contact, we use the Carnahan-Starling formula
\begin{equation}
4 \rho g(2a) 
 = \frac{1 + \rho + \rho^2 - \rho^3}{(1-\rho)^3} - 1 
 = 4 \rho \frac{1-\rho/2}{(1-\rho)^3}~,
\end{equation}
in 3D from Ref.\ \cite{hansen86}.
Initially, the normalized energies are $T=1$ 
and $R=0$ for all data presented here. 

\subsubsection*{Constant $r=0.99$, variable $\beta$ and $\rho$}

\begin{figure}[htb]
\begin{center}
\epsfig{file=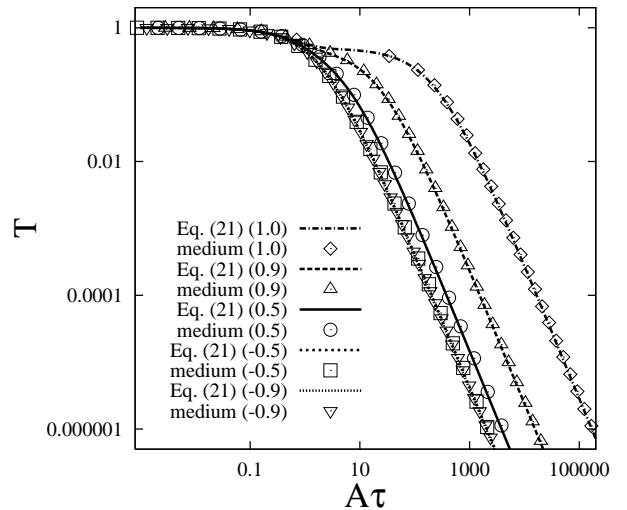,height=8.5cm,angle=-90}
\end{center}
\caption{
$T$ as function of rescaled time $A\tau$ in 3D. The symbols
correspond to simulations with $N=1331$, $\rho=0.087$ (medium),
$r=0.99$, and different $\beta$ as given in brackets in the inset.
The curves represent numerical solutions of Eqs.\
(\protect{\ref{eq:TR2}}) with the three-dimensional
constants from Eqs.\ (\protect{\ref{eq:const3d}}).
}
\label{fig:fig4}
\end{figure}

In Fig.\ \ref{fig:fig4} we plot $T$ against normalized time $A \tau$
for $r=0.99$ and various values of the tangential restitution $\beta$.
We observe a very similar picture as in 2D and, again, reasonable
agreement between theory and simulation over many orders of magnitude
in time. For $\beta<0.5$ most of the dependence on $\beta$ is taken
into account by our scaling,  $\tau\rightarrow A \tau$, so that the
scaled data almost collapse for $\beta < 0.5$.     

\begin{figure}[htb]
\begin{center}
\epsfig{file=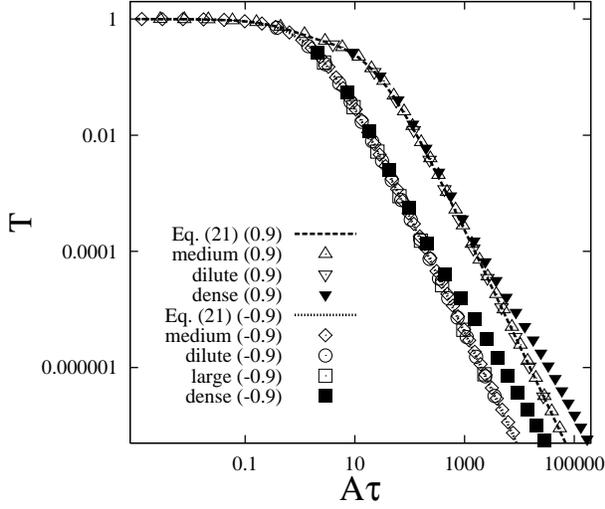,height=8.5cm,angle=-90} 
\end{center}
\caption{
$T$ as function of rescaled time $A\tau$ in 3D from simulations
with $r=0.99$, and $\beta=-0.9$ or $\beta=+0.9$ as given in 
brackets. Different symbols
correspond to simulations with $N=1331$, $\rho=0.087$ (medium), 
$N=4096$, $\rho=0.0021$ (dilute), $N=68921$, $\rho=0.0023$
(large), and $N=54872$, $\rho=0.23$ (dense). 
The curves are numerical solutions of Eqs.\
(\protect{\ref{eq:TR2}}) with the three-dimensional
constants from Eqs.\ (\protect{\ref{eq:const3d}}).
}
\label{fig:fig4x}
\end{figure}

In Fig.\ \ref{fig:fig4x} we compare  simulations of different
systems with the numerical solution of Eqs.\ \ref{eq:TR2}. Only the
{\em dense} simulations deviate from the theoretical result.  The
scaling of time with $A \tau$ is rather successful for small and
moderately dilute systems; in dense systems a deviation from the theory
occurs for large $A \tau$.

\begin{figure}[htb]
\begin{center}
\epsfig{file=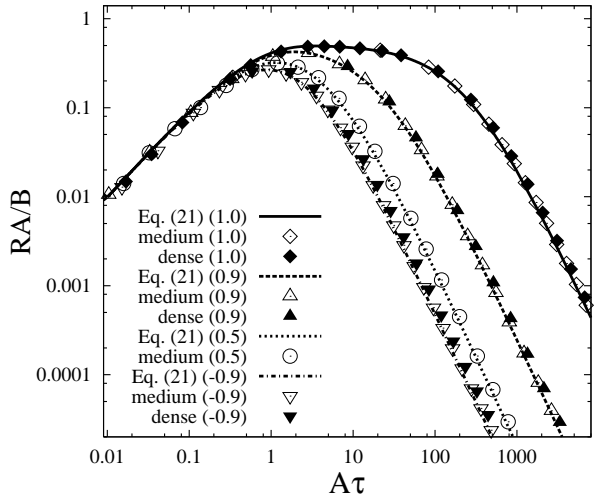,height=8.5cm,angle=-90}
\end{center}
\caption{
$RA/B$ as function of rescaled time $A\tau$. The data
are selected situations from Fig.\ \protect{\ref{fig:fig4}}.
The curves represent the numerical solutions of Eqs.\
(\protect{\ref{eq:TR2}}).
}
\label{fig:fig5}
\end{figure}

In Fig.\ \ref{fig:fig5} we plot $RA/B$ versus
normalized time $A \tau$ for  {\em medium} and 
{\em dense} systems.
As in 2D, we find that $RA/B$ increases proportional to
$A \tau$ for small times ($A \tau < 0.1$), reflecting the
activation of initially `cold' rotational degrees of freedom due
to collisions. This feature as well as the full time dependence is
well reproduced by our theoretical analysis.

\subsubsection*{Constant $\beta =-0.9$, variable $r$ and $\rho$}

\begin{figure}[htb]
\begin{center}
\epsfig{file=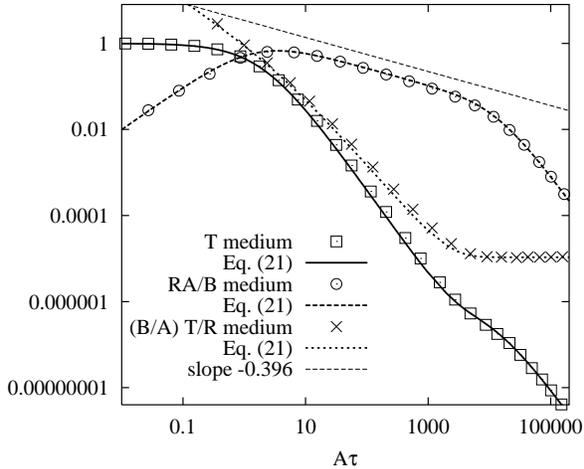,height=8.5cm,angle=-90}
\end{center}
\caption{
$T$, $RA/B$, and $(B/A) T/R$ for 3D simulations with 
$\beta = -0.9$ and $r=0.6$ in the medium (open symbols)
and in the dense (solid symbols) system from 
Fig.\ \protect{\ref{fig:fig4}}. The thick lines give the
solution of Eqs.\ (\protect{\ref{eq:TR2}}).
}
\label{fig:fig6}
\end{figure}

\begin{figure}[htb]
\begin{center}
\epsfig{file=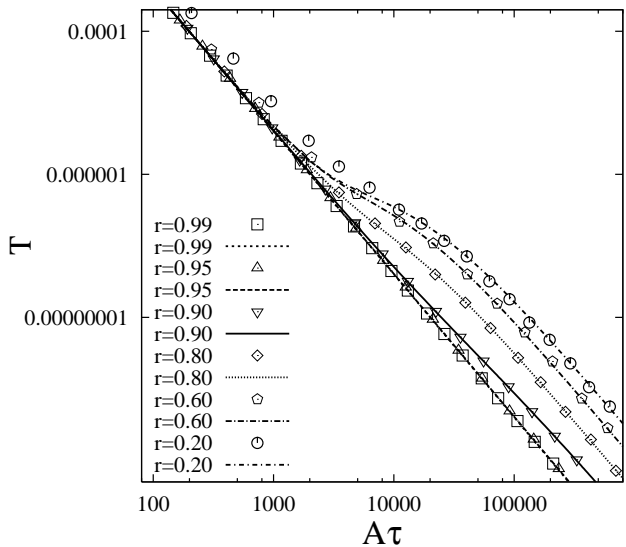,height=8.5cm,angle=-90}
\end{center}
\caption{
$T$ as function of rescaled time $A\tau$. The density is
$\rho=0.087$, $\beta=-0.9$ and the restitution coefficient
$r$ is given in the insert.
The curves represent again the numerical solutions of Eqs.\
(\protect{\ref{eq:TR2}}).
}
\label{fig:fig4r}
\end{figure}

In Fig.\ \ref{fig:fig6} we
present $T$, $RA/B$ and the ratio of $T$ and
$RA/B$ as a function of scaled time $A \tau$ for 
$r=0.6$ and $\beta = -0.9$, where interesting structure is observed.
The open symbols correspond to the {\it medium} 
system with $\rho=0.087$, $N=1331$. The data
are in good agreement with the theoretical curves, whilst we obtain
substantial differences between theory and simulation in the case 
of the dense system, due to the density instability [not shown here]. 
For the medium system the loss of energy during
collisions is predominantly due to normal restitution and only after
the translational energy has decayed to a very small value ($T<
10^{-5}$) does one observe the energy loss due to friction. The two
regimes can be discussed analytically with help of
Eqs.\ (\ref{eq:TR2}). For intermediate times, when the translational
energy is still appreciable, the equations can be simplified
for almost smooth spheres, i.e. ($\beta \approx -1$):
\begin{align}
\label{eq:smoothT}
\frac{d}{d \tau} T & = -A T^{3/2} \\
\frac{d}{d \tau} R & = - C T^{1/2}R~.
\label{eq:smooth}
\end{align}
We have neglected terms of ${\cal O}((1+\beta)^2)$ and approximate 
$A \approx (1-r^2)/4$ and $C \approx 5(1+\beta)/14$. The solution for
$T$ is that of smooth spheres, decaying like $T(\tau) \approx
(A\tau/2)^{-2}$ for large $\tau$. Substituting this result into the
equation for $R$, we find $R(\tau)/R(\tau_0)=(\tau/\tau_0)^{-\alpha}$
with $\alpha=2C/A$. Here $\tau_0$ is some intermediate timescale,
larger than the time for the initial increase of $R$, but smaller than the
timescale to reach the true asymptotic state. The above algebraic
decay is shown in Fig.\ \ref{fig:fig6} as a straight dashed line
with $\alpha \approx 0.396$. Once
the translational energy has decayed to a very small value as compared
to the rotational energy, all terms in the differential equations for 
$R$ and $T$ are equally important. We then observe a crossover from a  
$\tau^{-\alpha}$ to a $\tau^{-2}$ decay of the rotational energy. This
true asymptotic state is characterized by a constant ratio
$T/R$ and has been discussed above.
The crossover between the two regimes shows up as a parallel shift for
$T$ [see Figs.\ \ref{fig:fig6} and \ref{fig:fig4r}], because the translational 
energy decays like $\tau^{-2}$ in both regimes, but with a different prefactor.

When $r$ is increased to a
value close to unity, i.e. the elastic case, the intermediate time regime
disappears, because normal and tangential restitution are equally important.
This is demonstrated in Fig.\ \ref{fig:fig4r}, where we show $T$ 
plotted against the normalized time $A \tau$ for {\em medium} density
$\rho=0.087$, $\beta=-0.9$ and different values of $r$,
as given in the legend.

In the intermediate time
regime all curves follow the decay of smooth spheres 
[see Eq.(\ref{eq:smoothT})], which is
independent of $r$, because we use scaled time $A\tau $. 
In the true
asymptotic regime, all curves have the same slope with however an
axial section, which increases with decreasing $r$. 
Without scaling $\tau$ with $A$  the 
axial section decreases with decreasing $r$ reflecting the more
efficient dissipation of energy for smaller $r$.
The agreement between theory and
simulations is quite good for values of $r$ as low as $r=0.6$,
and even for $r=0.2$ only the crossover regime is not captured
by theory.

\section{Summary and Discussion}
\label{discuss}
Homogeneous cooling of colliding inelastic rough spheres has been
investigated with numerical simulations and an approximate kinetic
theory in two and three dimensions as well. 
We have confirmed that surface roughness is an important 
characteristic of the grains, in so far as it determines the decay of
translational energy, i.e. the rate of cooling. If energy loss due to
small normal and tangential restitution are comparable, then one observes an
initial linear change of translational and rotational energy, followed
by a crossover to the asymptotic regime, where both functions decay
like $t^{-2}$. This regime is characterized by a constant ratio
$T/R$, whose value depends on ${\it both}$, $r$ and $\beta$. The
dependence on $\beta$ is nonmonotonic, the ratio being smallest for
$\beta =\pm 1$.
This nonmonotonic dependence on $\beta$ also holds for the crossover
time, which is longest for $\beta =\pm 1$. If the coefficients of
normal and tangential restitution are such that energy is lost mainly
due to normal restitution, then we observe an intermediate time regime,
in between the initial linear change and the true asymptotic behaviour
with constant ratio $T/R$. This intermediate regime is also
characterized by an ${\it algebraic}$ decay of translational and
rotational energy: Translational 
energy decays like for smooth spheres ($t^{-2}$), whereas  
rotational energy decays with an exponent that depends
continuously on $r$ and $\beta$.

The theoretical approach is based on the assumption of a generalised
homogeneous cooling state: The $N$-particle distribution is assumed to
depend on time only via the average energies of translation and
rotation. Based on this assumption we computed the time decay of
translational energy and rotational energy without further
approximations. Good agreement with numerical simulations was found for
a large range of timescales and parameter sets, provided no density
instability builds up. The initial linear change,
the asymptotic $t^{-2}$ behaviour, the crossover in between as well as
the intermediate algebraic decay for almost smooth spheres - all these
features are accurately reproduced by our theoretical ansatz. 
These findings certainly  support the assumption of a homogeneous cooling
state and suggest to expand
around the HCS state to study deviations from homogeneous cooling.

\section*{Acknowledgements}

S.L. thanks the SFB 382 (A6), and M.H. thanks
the Land Niedersachsen for financial support.
S.M. gratefully acknowledges 
the support of the Alexander von Humboldt-Stiftung.

\begin{appendix}
\section{Detailed calculation in 2D}

In this appendix we explain, as an example, the main steps to
calculate $\left\langle i {\cal L}_+ E_{\rm tr}\right\rangle_{\rm HCS}$ of
Eq.\ (\ref{eq:hcs2}) in 2D.  The expectation value is calculated with the
$N$-particle distribution function, properly normalised
\begin{multline}
 \rho_{\rm HCS}(\Gamma;t) = \frac{1}{V^N}\left(\frac{m}{2\pi T_{\rm tr}(t)}
                                         \right)^{N}
                   \left(\frac{I}{2\pi T_{\rm rot}(t)}\right)^{N/2} \times \\ 
  \exp\left[-\sum_{\nu=1}^N \left( \frac{m}{2 T_{\rm tr}(t)} 
  \mbox{\boldmath$v$}_\nu^2+ \frac{I}{2T_{\rm rot}(t)}\omega_\nu^2\right) 
      \right]~.
\end{multline}
The angular velocity is a scalar in two dimensions, but a vector in
more than two dimensions.  Free streaming does not change the energy,
so we have to take into account only the collison operator 
$i{\cal L}_+'$ and we keep the abbreviation for $\Gamma$
from Eq.\ (\ref{eq:dgamma}) so that
\begin{multline}
\left\langle i{\cal L}_+' E_{\rm tr}\right\rangle_{\rm HCS} = \\ 
-\frac{1}{2}\sum_{\alpha\ne \beta}
\int d\Gamma \,
\rho_{\rm HCS}(\Gamma;t) 
{\cal C}_+(\alpha\beta) \frac{1}{2N} \sum_{\nu=1}^{N}m \mbox{\boldmath$v$}_\nu^2=\\
-\frac{1}{2N}\sum_{\alpha\ne \beta}
\int d\Gamma \,
\rho_{\rm HCS}(\Gamma;t) 
{\cal C}_+(\alpha\beta) \frac{m}{2}
\left(\mbox{\boldmath$v$}_\alpha^2+\mbox{\boldmath$v$}_\beta^2\right)~.
\end{multline}
The binary collision operator ${\cal C}_+(\alpha\beta)$ gives a contribution
only, if either $\nu=\alpha$ or if $\nu=\beta$. Next, we introduce two
$\delta$-functions,
\begin{multline}\label{eq:a3}
\left\langle i{\cal L}_+' E_{\rm tr}\right\rangle_{\rm HCS} =    \\          
 - \frac{1}{2N}\sum_{\alpha\ne \beta}
\int d \Gamma
\int d \mbox{\boldmath$R$}_1 d\mbox{\boldmath$R$}_2 
\delta(\mbox{\boldmath$R$}_1-\mbox{\boldmath$r$}_\alpha)
\delta(\mbox{\boldmath$R$}_2-\mbox{\boldmath$r$}_\beta)          \\
\rho_{\rm HCS}(\Gamma;t) 
{\cal C}_+(\alpha\beta) \frac{m}{2} 
\left(\mbox{\boldmath$v$}_\beta^2+\mbox{\boldmath$v$}_\alpha^2\right)~,
\end{multline} 
which allows us to replace $\mbox{\boldmath$r$}_\alpha$ by
$\mbox{\boldmath$R$}_1$ and $\mbox{\boldmath$r$}_\beta$ by
$\mbox{\boldmath$R$}_2$ in ${\cal C}_+(\alpha\beta)$. Integration over all
${\mbox{\boldmath$r$}_{\mu}}$ of the respective part of Eq. (\ref{eq:a3})
can then be performed and yields a
factor
\begin{multline}
  \int \prod_{\mu=1}^{N} d\mbox{\boldmath$r$}_\mu\prod_{\mu<\kappa}
  \Theta\left(|\mbox{\boldmath$r$}_{\mu\kappa}|-2a \right)
  \delta(\mbox{\boldmath$R$}_1-\mbox{\boldmath$r$}_\alpha)
  \delta(\mbox{\boldmath$R$}_2-\mbox{\boldmath$r$}_\beta) =     \\
  V^{N-2}g(\mbox{\boldmath$R$}_{12})
\end{multline}
The pair correlation function $g(\mbox{\boldmath$R$}_{12})$ depends on
$|\mbox{\boldmath$R$}_{12}| = 
     |\mbox{\boldmath$R$}_{1}-\mbox{\boldmath$R$}_{2}|$ only. 
Similarly integration over all
velocities and angular velocities with index $\mu$ and 
$\alpha\ne \mu \ne \beta$ 
gives $1$ due to normalization. We can sum over $N(N-1)$
identical integrals and get
\begin{multline}
  \left\langle i{\cal L}_+' E_{\rm tr} \right\rangle_{\rm HCS} =        \\ 
  - \frac{(N-1)}{2 V^2} 
  \left(\frac{m}{2\pi  T_{\rm tr}(t)}\right)^{2}
  \frac{I}{2\pi T_{\rm rot}(t)} 
  \int d \omega_1 d\omega_2
  d \mbox{\boldmath$R$}_1 d\mbox{\boldmath$R$}_2 
  d \mbox{\boldmath$v$}_1 d\mbox{\boldmath$v$}_2                  \\
  \exp\left(
    -\frac{m}{2 T_{\rm tr}(t)} (\mbox{\boldmath$v$}_1^2+\mbox{\boldmath$v$}_2^2) -
    \frac{I}{2T_{\rm rot}(t)}(\omega_1^2+\omega_2^2 )  \right)            \\
  g(\mbox{\boldmath$r$}) 
  (\mbox{\boldmath$v$}_{12}\cdot\hat{\mbox{\boldmath$r$}})
  \Theta\left(- \mbox{\boldmath$v$}_{12}\cdot\hat{\mbox{\boldmath$r$}}\right)
  \delta\left(|\mbox{\boldmath$r$}|-2a \right) \Delta E_{\rm tr}~.
\end{multline}
The loss of translational energy of two colliding particles
is denoted by $\Delta E_{\rm tr}$ and given by
\begin{multline}
  \Delta E_{\rm tr}=  \frac{m}{2}
  \big[2\eta(\eta-1)(\mbox{\boldmath$v$}_{12}^2-(\mbox{\boldmath$v$}_{12}
  \cdot\hat{\mbox{\boldmath$r$}})^2)-       \\ 
  (1/2)(1-r^2)(\mbox{\boldmath$v$}_{12}\cdot\hat{\mbox{\boldmath$r$}})^2
  + 2 \eta^2 a^2 \Omega^2 \big]~.
\end{multline}
Here we use the abbreviations $\eta= \frac{q(1+\beta)}{2q+2} $,
$\Omega=(\omega_1+\omega_2)/\sqrt{2}$, and
$\mbox{\boldmath$R$}_1-\mbox{\boldmath$R$}_2=\mbox{\boldmath$r$}=r\hat{\mbox{\boldmath$r$}}$.
To perform the remaining integrations we substitute
\begin{alignat}{2}
                            \Omega & =\frac{1}{\sqrt{2}}(\omega_1+\omega_2),
\quad &                     \omega & =\frac{1}{\sqrt{2}}(\omega_1-\omega_2), \\
               \mbox{\boldmath$V$} & =\frac{1}{\sqrt{2}}(\mbox{\boldmath$v$}_1+\mbox{\boldmath$v$}_2),
\quad &        \mbox{\boldmath$v$} & =\frac{1}{\sqrt{2}}(\mbox{\boldmath$v$}_1-\mbox{\boldmath$v$}_2),\\
               \mbox{\boldmath$r$} & =\mbox{\boldmath$R$}_1-\mbox{\boldmath$R$}_2,
\quad &        \mbox{\boldmath$R$} & =\mbox{\boldmath$R$}_1,
\end{alignat}
The Jacobian determinant for the above transformation is $1$.
Integration over $\omega$, $\mbox{\boldmath$V$}$ and
$\mbox{\boldmath$R$}$ can be done, which all give the value $1$ due to
normalisation.  The resulting integral is
\begin{multline} \nonumber
  \left\langle i{\cal L}_+' E_{\rm tr}\right\rangle_{\rm HCS} = 
  - \frac{(N-1)m}{4\pi T_{tr}(t) V} \left(\frac{2I}{2\pi T_{rot}(t)}\right)^{1/2}  
  \int d \Omega d \mbox{\boldmath$r$} d\mbox{\boldmath$v$} \\
  \exp \left(
    -\frac{m\mbox{\boldmath$v$}^2}{2
      T_{\rm tr}(t)} - \frac{I\Omega^2}{2T_{\rm rot}(t)}
  \right)   
  g(\mbox{\boldmath$r$}) (\mbox{\boldmath$v$}\cdot\hat{\mbox{\boldmath$r$}})
  \Theta \left(
    -\mbox{\boldmath$v$}\cdot\hat{\mbox{\boldmath$r$}}
  \right)
  \delta\left(|\mbox{\boldmath$r$}|-2a \right)\\
 \frac{m}{2} \big [
  2\eta(\eta-1)(\mbox{\boldmath$v$}^2-
  (\mbox{\boldmath$v$}\cdot\hat{\mbox{\boldmath$r$}})^2)\\
  -(1/2)(1-r^2)(\mbox{\boldmath$v$}\cdot\hat{\mbox{\boldmath$r$}})^2
  +2\eta^2 a^2 \Omega^2 
  \big ]~.
\end{multline}
The integration over $|\mbox{\boldmath$r$}|$ yields $2ag(2a)$. Choosing e.g.
$\hat{\mbox{\boldmath$r$}}$ to point along the $x$-axis, the integrals
over linear and angular velocities can easily be done as moments of a
Gaussian distribution. The result is independent of
$\hat{\mbox{\boldmath$r$}}$, so that the integration over
$\hat{\mbox{\boldmath$r$}}$ gives $2\pi$.  Finally we assume that
$1\ll N$, approximate $N\approx(N-1)$ and obtain the result of
Eq.\ (\ref{eq:hcs2})

\section{ED Algorithm}
Simple ED algorithms update the whole system after each event, a
method which is straightforward, but inefficient for large numbers of
particles. In Ref. \cite{lubachevsky91} an ED algorithm was introduced
which updates only those two particles which were involved in the last
collision. For this a double buffering data structure is implemented,
which contains the `old' status and the `new' status, each consisting
of: time of event, position, velocities, and event-partner. When a
collision occurs, the `old' and `new' status of the participating
particles are exchanged. Thus, the former `new' status becomes the
actual `old' one, while the former `old' status becomes the `new' one
and is free for future calculations. This seemingly complicated
exchange from information is carried out extremely simple and fast by
only exchanging the pointers to the `new' and `old' status
respectively. The `old' status of particle $i$ has to be kept in
memory, in order to calculate the time of the next contact, $t_{ij}$,
of particle $i$ with any other object $j$ which can change its status 
due to a collision with yet another particle. During the simulation this
may be neccessary several times so that the predicted `new' status
has to be modified. An object $j$ is either a particle ($j$ = 1, ...,
$i-1$, $i+1$, ..., $N$) or a cell-wall ($j$ = $N+1$, ... ). The
minimum of all $t_{ij}$ is stored in the `new' status of particle $i$,
together with the corresponding partner $j$. Depending on the
implementation, also positions and velocities after the collision can
be calculated.  This would be a waste of computer time, since before
the time $t_{ij}$, the predicted partners $i$ and $j$ might be
involved into several collisions with other particles, so that we
apply a delayed update scheme \cite{lubachevsky91}. The minimum
times of event, i.e. the times which indicate the next event for a
certain particle, are stored in an ordered heap tree, such that the
next event is found at the top of the heap with computational effort
of ${\cal O}(1)$; changing the position of one particle in the tree
from the top to a new position needs ${\cal O}(\log N)$ operations.
The search for possible collision partners is accelerated 
by the use of a standard linked-cell data structure and consumes ${\cal
O}(1)$ of numerical ressources. In total, this results in numerical 
effort of ${\cal O}(N \log N)$ for $N$ particles. For a detailed description 
of the algorithm see Ref.\ \cite{lubachevsky91}.  
Using all these algorithmic tricks, we are able
to simulate up to $10^5$ particles within reasonable time on a small
workstation (IBM P43/133MHz) \cite{luding98}.  The particle number is
limited by RAM-size (64MB) rather than CPU-power.
 
\end{appendix}

\end{document}